\documentclass{article}
\usepackage{arxiv}

\usepackage[utf8]{inputenc} % allow utf-8 input
\usepackage[T1]{fontenc}    % use 8-bit T1 fonts
\usepackage[hidelinks]{hyperref}   % hyperlinks
\usepackage{url}            % simple URL typesetting
\usepackage{booktabs}       % professional-quality tables
\usepackage{amsfonts}       % blackboard math symbols
\usepackage{nicefrac}       % compact symbols for 1/2, etc.
\usepackage{microtype}      % microtypography
\usepackage{graphicx}
\usepackage{amsmath}
\usepackage[table]{xcolor}
\usepackage[warn]{textcomp}
\usepackage{epsfig}
\usepackage{verbatim}
\usepackage{multirow}
\usepackage{txfonts}

\title{TopologyGAN: Topology Optimization Using Generative Adversarial Networks Based on Physical Fields Over the Initial Domain}

\author{
  Zhenguo Nie \\
  Department of Mechanical Engineering\\
  Carnegie Mellon University\\
  Pittsburgh, PA 15213\\
  \texttt{zhenguon@andrew.cmu.edu} \\
   \And
  Tong Lin \\
  Department of Mechanical Engineering\\
  Carnegie Mellon University\\
  Pittsburgh, PA 15213\\
  \texttt{tongl1@andrew.cmu.edu} \\
   \And
  Haoliang Jiang \\
  School of Computer Science\\
  Georgia Institute of Technology\\
  Atlanta, GA 30332\\
  \texttt{hjiang321@gatech.edu} \\
  \And
  Levent Burak Kara\thanks{Address all correspondence to this author.} \\
  Department of Mechanical Engineering\\
  Carnegie Mellon University\\
  Pittsburgh, PA 15213\\
  \texttt{lkara@cmu.edu} \\
}
\begin{document}
\maketitle

%%%%%%%%%%%%%%%%%%%%%%%%%%%%%%%%%%%%%%%%%%%%%%%%%%%%%%%%%%%%%%%%%%%%%%
\begin{abstract}
In topology optimization using deep learning, load and boundary conditions represented as vectors or sparse matrices often miss the opportunity to encode a rich view of the design problem, leading to less than ideal generalization results. We propose a new data-driven topology optimization model called TopologyGAN that takes advantage of various physical fields computed on the original, unoptimized material domain, as inputs to the generator of a conditional generative adversarial network (cGAN). Compared to a baseline cGAN, TopologyGAN achieves a nearly $3\times$ reduction in the mean squared error and a $2.5\times$ reduction in the mean absolute error on test problems involving previously unseen boundary conditions. Built on several existing network models, we also introduce a hybrid network called U-SE(Squeeze-and-Excitation)-ResNet for the generator that further increases the overall accuracy. We publicly share our full implementation and trained network.

%The full implementation (based on Tensorflow) and the trained networks are available at \url{https://github.com/zhenguonie/TopologyGAN}.


\end{abstract}

%%%%%%%%%%%%%%%%%%%%%%%%%%%%%%%%%%%%%%%%%%%%%%%%%%%%%%%%%%%%%%%%%%%%%%
\section{Introduction}
\begin{figure}[!ht]
    \centering
    \includegraphics[width=0.5\linewidth]{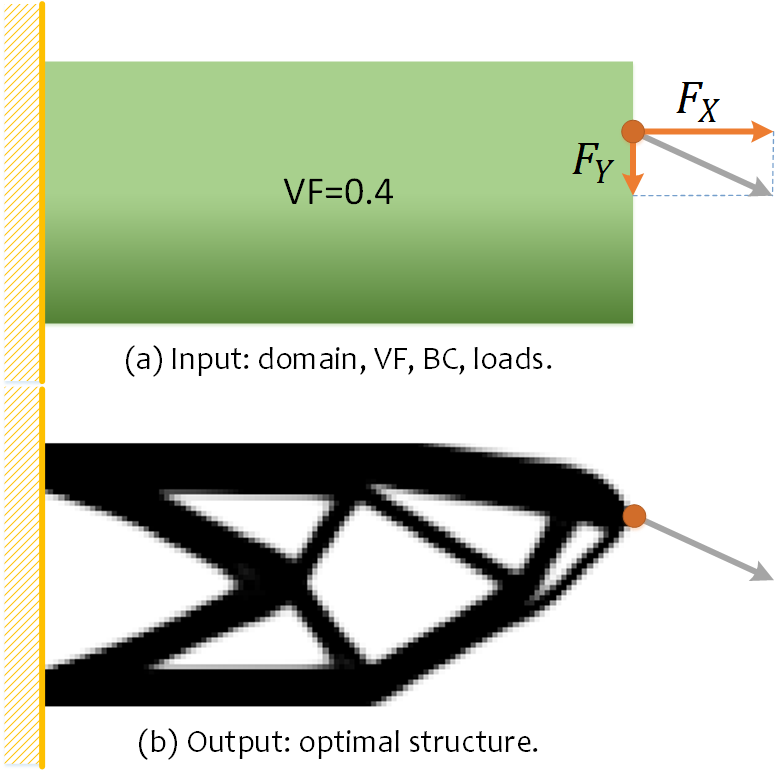}
    \caption{2D topology optimization.}
    \label{fig:topology_optimization}
\end{figure}

Topology optimization of solid structures involves generating optimized shapes by minimizing an objective function such as compliance or mass within a material domain, subject to a set of displacement and load boundary conditions (Figure \ref{fig:topology_optimization}). With rapid advances in additive manufacturing and the associated design tools, topology optimization is becoming increasingly more prevalent as it allows optimized structures to be designed automatically. Existing methods include the density based approaches such as the Solid Isotropic Material with Penalization (SIMP) method \cite{bendsoe1988generating,suzuki1991homogenization,bendsoe2009topology,diaz1992shape,bendsoe1989optimal,bendsoe1999material,sigmund2007morphology,biyikli2015proportional,nie2019optimization,suresh2010199}, grid based approaches \cite{xie1993simple,chu1997various,querin1998evolutionary,young19993d,reynolds1999reverse,liu2000metamorphic}, moving boundary based approaches \cite{eschenauer1994bubble,eschenauer1997topology,allaire2002level,allaire2004structural,allaire2005structural,wang2003level, challis2009level, suresh2013stress,ulu2017lightweight}, and load-path based approaches \cite{kelly1995procedure,kelly2001load,kelly2011interpreting}. Although significant efforts have been made to improve solution efficiency \cite{sigmund2013topology}, topology optimization methods remain to be computationally demanding and are not readily suited to be used inside other design optimization modules such as layout or configuration design tools. 

In recent years, new data-driven methods for topology optimization have been proposed to accelerate the process. Deep learning methods have shown promise in efficiently producing near optimal results with respect to shape similarity as well as compliance with negligible run-time cost \cite{guo2018indirect,yu2019deep,rawat2019application,sharpe2019topology,ulu2016data}. Theory-guided machine learning methods use domain-specific theories to establish the mapping between the design variables and the external boundary conditions \cite{cang2019one,lei2019machine,kumar2019density}. However a significant challenge in topology optimization is learning an accurate and generalizable mapping from the boundary conditions to the resulting optimal structure. As such, approaches that involve  establishing this map directly often have to severely restrict the displacement and external load configurations, as the results are difficult to generalize to novel boundary conditions.

As one step toward addressing this issue, we propose a new deep learning based generative model called TopologyGAN for topology optimization. TopologyGAN is based on conditional Generative Adversarial Networks (cGAN). The main hypothesis we pursue is that rather than trying to map the boundary conditions to the resulting optimal shapes directly, various physical fields computed on the initial, unoptimized domain subject to the prescribed boundary conditions may hold useful information that allows the network to learn more accurate maps. This approach has been motivated by our observation that in deep learning based approaches, the input displacement and load boundary conditions are often represented as sparse vectors or matrices\footnote{Unless, of course, they are fixed, in which case the network cannot account for variations in these conditions.}. This sparsity prevents the network from exploiting useful spatial variations and physical phenomena that occur within the material domain. By contrast, in this work, we propose to take advantage of such variations.

To this end, in TopologyGAN (\textit{i}) the input channels of the generator are related to the physical fields (non-sparse matrices) computed in the initial, unoptimized domain such as the von Mises stress fields, strain energy fields, and displacement fields rather than the original boundary conditions alone, and (\textit{ii}) the condition for the discriminator involve both the physical fields and the initial inputs. We use ground truth data generated by the SIMP method, although our approach is applicable to ground truth data obtained from other topology optimization methods.

%For deep learning based approaches, the input load and the displacement boundary conditions are usually represented as sparse matrices. Due to the high sparsity, the convolution operation is not able to extract sufficient information and therefore, such a mechanism doesn't result in good generalizability. To solve this problem, our approach incorporates the dense matrices, which provide further information related to the energy distribution of the target structure, into the model inputs. Such information is the physical fields of the original domain that are derived from FEM on the initial domain and is fed into the model as a separate channel in addition to the BC, load, and VF matrices. 

Based on this formulation and the model selection studies we conduct, we find that the von Mises stress field and the strain energy density fields are the most useful channels of information to augment with the original displacement and load boundary conditions. Our results show that compared to a baseline cGAN model that does not take advantage of such fields, TopologyGAN achieves lower \emph{test} errors than that of the \emph{training} error of the baseline method on previously unseen boundary conditions. These results are encouraging in that they may prove useful for other researchers who may wish to explore the use of such fields in similar topology optimization scenarios. 

We publicly share our full implementation and trained network (\url{https://github.com/zhenguonie/TopologyGAN}). The entire data set used in this work is available freely upon request. 

Our main contributions are:
\begin{description}
  \item[$\bullet$] A new method, TopologyGAN, for topology optimization using deep learning models.
    \item[$\bullet$] A new design of the input matrices involving the initial physical fields. This input complements the original problem input matrices.
  \item[$\bullet$] A hybrid neural network architecture, namely U-SE-ResNet, as the generator for TopologyGAN.
\end{description}

%%%%%%%%%%%%%%%%%%%%%%%%%%%%%%%%%%%%%%%%%%%%%%%%%%%%%%%
\section{Related Work}
Our review focuses on studies that highlight topology optimization, deep learning for topology optimization, generative adversarial networks (GAN), and two network architectures closely related to our work.

\subsection{Topology Optimization and SIMP}
Topology optimization seeks an optimal subset $\Omega_{mat} \subset \Omega$, where $\Omega$ is the design domain. To formulate this problem, an objective function $f(y)$ to be minimized is defined as in Eq.(\ref{equ:objective}), in which $y$ denotes the structural design (material distribution)\footnote{We use $y$ for material distribution for consistency with the cGAN output presented in the following sections.} and $h(y)$ is a resulting physical outcome such as stress, strain or displacement.

\begin{equation}\label{equ:objective}
f(y)=\left\{
\begin{aligned}
\rm \mathop{min} \limits_{y} & & f(y,h(y)) \\
\texttt{s.t.} & & \left\{
                \begin{aligned} \rm{design\ constraint\ on}\ \it{y}\\
                \rm{state\ constraint\ on}\ \it{h(y)}\\
                \rm{equilibrium\ constraint}
                \end{aligned}
                \right.
\end{aligned}
\right.
\end{equation}

In this work, we use the density-based SIMP method (\cite{bendsoe1988generating, rozvany1992generalized}), which is widely implemented in many commercial design software \cite{eschenauer2001topology,plocher2019review}. The SIMP method discretizes the design domain into a grid of finite elements called isotropic solid microstructures. It models the material density $y_e$ to vary between zero (void) and one (full solid). This representation allows the assignment of intermediate densities to the elements. The Young's modulus $E_e$ for each grid element $e$ is given as:

\begin{equation}\label{equ:youngs}
E_e(y_e) = E_{min}+{y}^{p}_{e}(E-E_{min})
\end{equation}

\noindent where $E$ is the material stiffness, $E_{min}$ is an infinitesimal stiffness and $p$ is a penalization factor to favor binary outputs avoiding intermediate densities. The optimization works toward minimizing the compliance $C(y)$ ( \cite{ambrosio1993optimal, petersson1999some}) as follows:
\begin{equation}\label{equ:simp}
\left.
    \begin{aligned}
        \mathop{\text{min}} \limits_{y}: & \quad C(y) = \mathbf{U}^T \mathbf{KU}=\sum_{e=1}^{N} (y_e)^p \mathbf{u}_e^T \mathbf{k}_e \mathbf{u}_e \\
                        \texttt{s.t.}: & \quad \tfrac{V(y)}{V_0}=\textrm{VF} \\
                                     : & \quad \mathbf{KU=F} \\
                                     : & \quad 0 \leq y_e \leq 1
    \end{aligned}
\right\}
\end{equation}

\noindent where $y$ is the density-based structural design tensor, \textbf{U} and \textbf{F} are the global displacement and force vectors, \textbf{K} is the stiffness matrix, $\mathbf{u}_e$ is the elemental displacement vector, $\mathbf{k}_e$ is the elemental stiffness matrix, and $N$ is the number of total elements.

While the above existing methods can provide optimized solutions, our work aims to accelerate the iterative nature of these solvers using a data-driven approach. 

\subsection{Deep Learning for Topology Optimization}
With recent advances in computer hardware, deep neural networks have been widely applied in various fields, including autonomous vehicles, robotics, medical diagnosis, bio-medicine, material design, machine health monitoring, mechanical design, and manufacturing. Deep neural networks have proven to be effective at learning complex mappings between problem input variables and constraints, and target design objectives. Supervised machine learning techniques have proven to be effective for engineering design exploration and optimization, and for mapping out feasible regions of the design space \cite{sharpe2019comparative}. Guo et al. \cite{guo2018indirect} propose a data-driven design representation where an augmented variational autoencoder is used to encode 2D topologies into a lower-dimensional latent space and to decode samples from this space back into 2D topologies. Oh et al. \cite{oh2019deep} propose a deep generative adversarial framework capable of generating numerous design alternatives that are both aesthetic and optimized for engineering performance. In 3D topology optimization, Rawat and Shen \cite{rawat2019application} integrate Wasserstein GAN and a convolutional neural network (CNN) into a coupled model to generate new 3D structures using limited data samples. To speed up convergence in SIMP based topology optimization for thermal conduction, Lin et al. \cite{lin2018investigation} introduce a deep learning approach using U-Nets. Through deep learning, only the early results obtained through SIMP are fed into the network to directly produce the final outputs. Sosnovik and Oseledets \cite{sosnovik2019neural} use CNNs to accelerate topology optimization from two halfway grayscale images to the final binary image generated through SIMP. However, these networks focus either on latent candidate generation or accelerated optimization and do not establish an end-to-end mapping from the boundary conditions to the optimized topologies.

To realize an end-to-end topology optimization from prescribed boundary conditions, Yu et al. \cite{yu2019deep} propose a CNN-based encoder-decoder for the generation of low-resolution structures, which are then passed through a super-resolution GAN to generate the final results. Sharpe and Seepersad \cite{sharpe2019topology} explore the use of cGANs as a means of generating a compact latent representation of structures resulting from topology optimization. However, only a few boundary conditions and optimization settings are considered. Extending data-driven topology optimization to novel displacement and external load conditions remains a major challenge. 

\subsection{Generative Adversarial Networks}
\begin{figure}[!ht]
    \centering
    \includegraphics[width=1.0\linewidth]{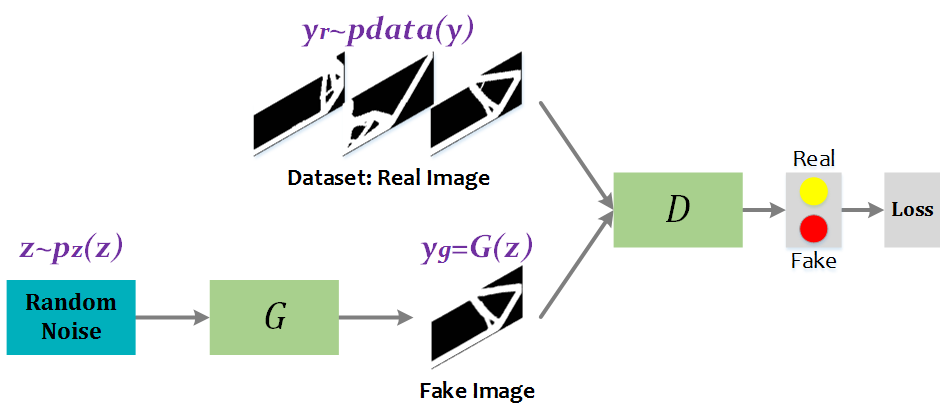}
    \caption{Schematic diagram of GAN.}
    \label{fig:vanilla_gan}
\end{figure}

Generative Adversarial Networks (GAN) \cite{goodfellow2014generative}, is a generative model formulated as a minimax two-player game between two models. It consists of: (1) A generator $\mathit{G}$ whose aim is to learn a generative density function that models the training data and, (2) a discriminator $\mathit{D}$ that aims to discern if an input sample is part of the original training set or a synthetic one generated by $\mathit{G}$. The structure of GAN is shown in Figure \ref{fig:vanilla_gan}. The input to $\mathit{G}$ is random noise $z$ sampled from a distribution $p_z(z)$. The output of $\mathit{G}$, $y_g=\mathit{G}(z)$, is a fake data. A real $y_r$ or a fake sample $y_g$ is then fed into $\mathit{D}$ to obtain an output $\mathit{D}(y)$, which generates a probability estimate of the true nature of $y$. In this case, $\mathit{D}$ is trained to maximize the probability of assigning the correct label to both the real samples $y_r$ and fake samples $y_g$. $\mathit{G}$ is simultaneously trained to minimize $\log(1-\mathit{D}(\mathit{G}(z)))$. The training loss functions $\mathcal{L}_\mathit{D}^{\mathit{GAN}}$ and $\mathcal{L}_\mathit{G}^{\mathit{GAN}}$ are:

%$\mathcal{L}_\mathit{G}$ and $\mathcal{L}_\mathit{G}$ can be expressed as

\begin{equation}\label{equ:gan_loss}
\mathop{\text{max}}\limits_{\mathit{D}} \mathcal{L}_\mathit{D}^{\mathit{GAN}}=\mathbb{E}_{y_r\sim p_{data}(y)}[\log\mathit{D}(y_r)]+\mathbb{E}_{z \sim p_z(z)}[\log(1-\mathit{D}(\mathit{G}(z)))]
\end{equation}

\begin{equation}\label{equ:gan_loss_g}
\mathop{\text{min}}\limits_{\mathit{G}} \mathcal{L}_\mathit{G}^{\mathit{GAN}}= \mathbb{E}_{z \sim p_z(z)}[\log(1-\mathit{D}(\mathit{G}(z)))]
\end{equation}

\begin{figure}[!ht]
    \centering
    \includegraphics[width=1.0\linewidth]{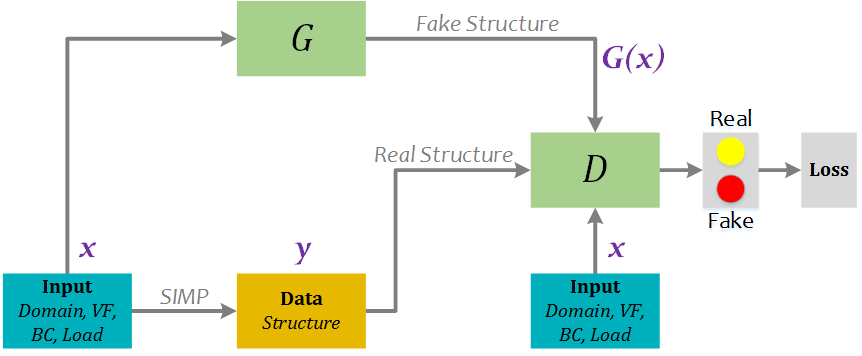}
    \caption{Baseline: cGAN for topology optimization.}
    \label{fig:cgan}
\end{figure}

\emph{Conditional GAN} Generative adversarial networks can be extended to a conditional model when the generator and the discriminator are conditioned on prescribed information \cite{mirza2014conditional}. A conditional GAN (cGAN) learns a mapping from an input $x$ to an output $y$, as shown in Figure \ref{fig:cgan}. The cGAN loss (\cite{isola2017image}) is:

\begin{equation}\label{equ:cgan_loss}
\begin{split}
\mathcal{L}_\mathit{G,D}^{\mathit{cGAN}}= \mathbb{E}_{(x,y) \sim p_{data}(x,y)}[\log\mathit{D}(x,y)]+ \\ \mathbb{E}_{x \sim p_{data}(x),z \sim p_z(z)}[\log(1-\mathit{D}(x,\mathit{G}(x,z)))]
\end{split}
\end{equation}

\begin{equation}\label{equ:cgan_loss_l1}
\mathcal{L}_{L1}(\mathit{G}) = \mathbb{E}_{x,y,z}[\left\|y-G(x,z)\right\|_1]
\end{equation}

\begin{equation}\label{equ:cgan_loss_final}
\mathit{G}^* = \text{arg}\mathop{\text{min}}\limits_{\mathit{G}}\mathop{\text{max}}\limits_{\mathit{D}} \ \mathcal{L}_\mathit{G,D}^{\mathit{cGAN}} + \lambda \mathcal{L}_{L1}(\mathit{G})
\end{equation}

\noindent where $\mathit{G}^*$ is the final optimized generative model. In this model, the inputs $x$ are composed of the original (full) domain, the desired volume fraction, the displacement boundary conditions, and the external loads. These are utilized as conditions that the generator has to attune to. The ground truth final optimized structure (real structure) $y$ is computed through the SIMP method and provided as input to the discriminator, alongside $x$ for training. 

\begin{comment}
\begin{figure}[!ht]
    \centering
    \includegraphics[width=0.9\linewidth]{figure/schematics_cgan.png}
    \caption{Schematic diagram of the conditional GAN \cite{isola2017image}.}
    \label{fig:schematics_cgan}
\end{figure}
\end{comment}

\subsection{U-Net and SE-ResNet}
\begin{figure}[!ht]
    \centering
    \includegraphics[width=0.5\linewidth]{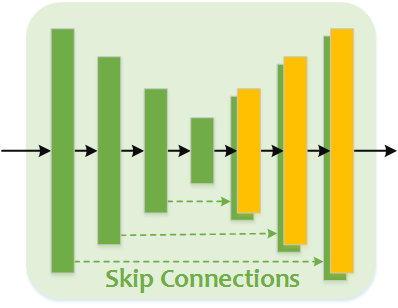}
    \caption{U-Net}
    \label{fig:unet}
\end{figure}

The U-Net architecture (\cite{long2015fully}) shown in Figure \ref{fig:unet} allows the network to propagate context information from the downsampling layers to the upsampling layers at various resolutions. To predict the output around the border of the image, the missing context is extrapolated by reflecting the input image. This tiling strategy is important when the network is applied to large images, as otherwise the resolution would be limited by the GPU capacity \cite{ronneberger2015u}.

\begin{figure}[!ht]
    \centering
    \includegraphics[width=1.0\linewidth]{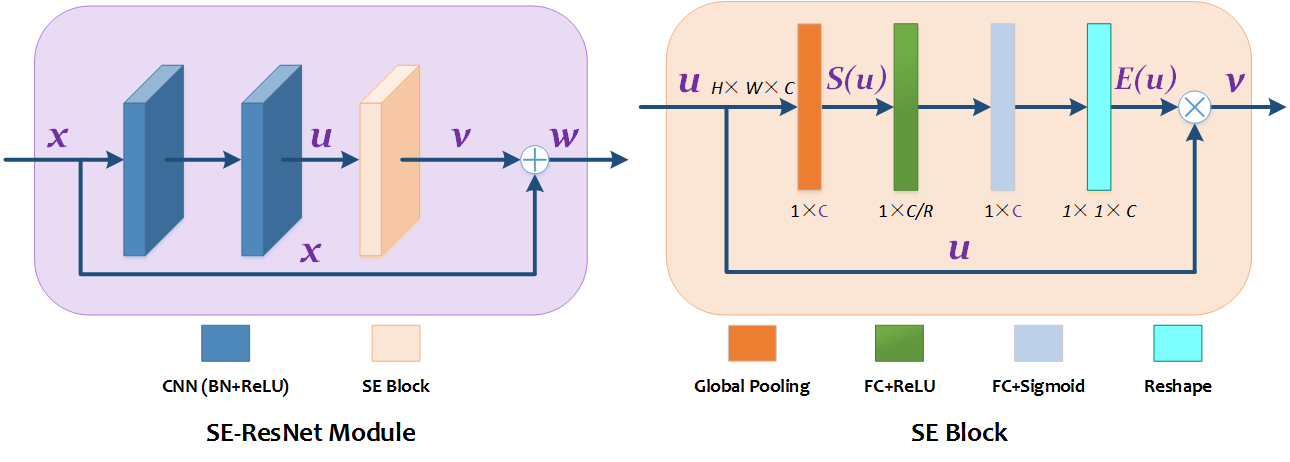}
    \caption{ResNet enhanced by SE block \cite{nie2020stress}}
    \label{fig:resnet_se}
\end{figure}

\emph{SE-ResNet} SE-ResNet is a CNN-based ResNet \cite{he2016deep} enhanced by SE-blocks \cite{hu2018squeeze} as shown in Figure \ref{fig:resnet_se}. Each SE-ResNet module contains two CNN blocks and one SE block \cite{nie2020stress}. The distinguishing feature of ResNet is the skip connection which simply performs identity mapping added to the output of the stacked layers. As such, ResNet can dynamically select the layer depth for the desired underlying mapping. The final output of the SE-ResNet module is computed by a feedforward neural network with a shortcut connection: $w=v+x$.

SE block is used in SE-ResNet to improve the representational capacity of the network by enabling it to perform dynamic channel-wise feature recalibration. The input data $u \in R^{H \times W \times C}$ is shrunk to $S(u) \in R^C$ through the global average-pooling layer. Then two fully connected layers are used to downsample and upsample the linear array $S(u)$. A reshape operation produces the excitation output $E(u)$ that has the same dimension as the initial input $u$. The final output of the SE block is a Hadamard product of $E(u)$ and $u$ through the element-wise matrix multiplication: $v = E(u) \otimes u$.

%%%%%%%%%%%%%%%%%%%%%%%%%%%%%%%%%%%%%%%%%%%%%%%%%%%%%%%
\section{Technical Approach}
Topology optimization using deep learning has difficulties in extending to previously unseen boundary conditions because of the high discreteness of the boundary conditions and the sparsity of the input matrices. The high sparsity of the input matrices leads to a high variance of the mapping function.

\begin{figure}[!ht]
    \centering
    \includegraphics[width=1.0\linewidth]{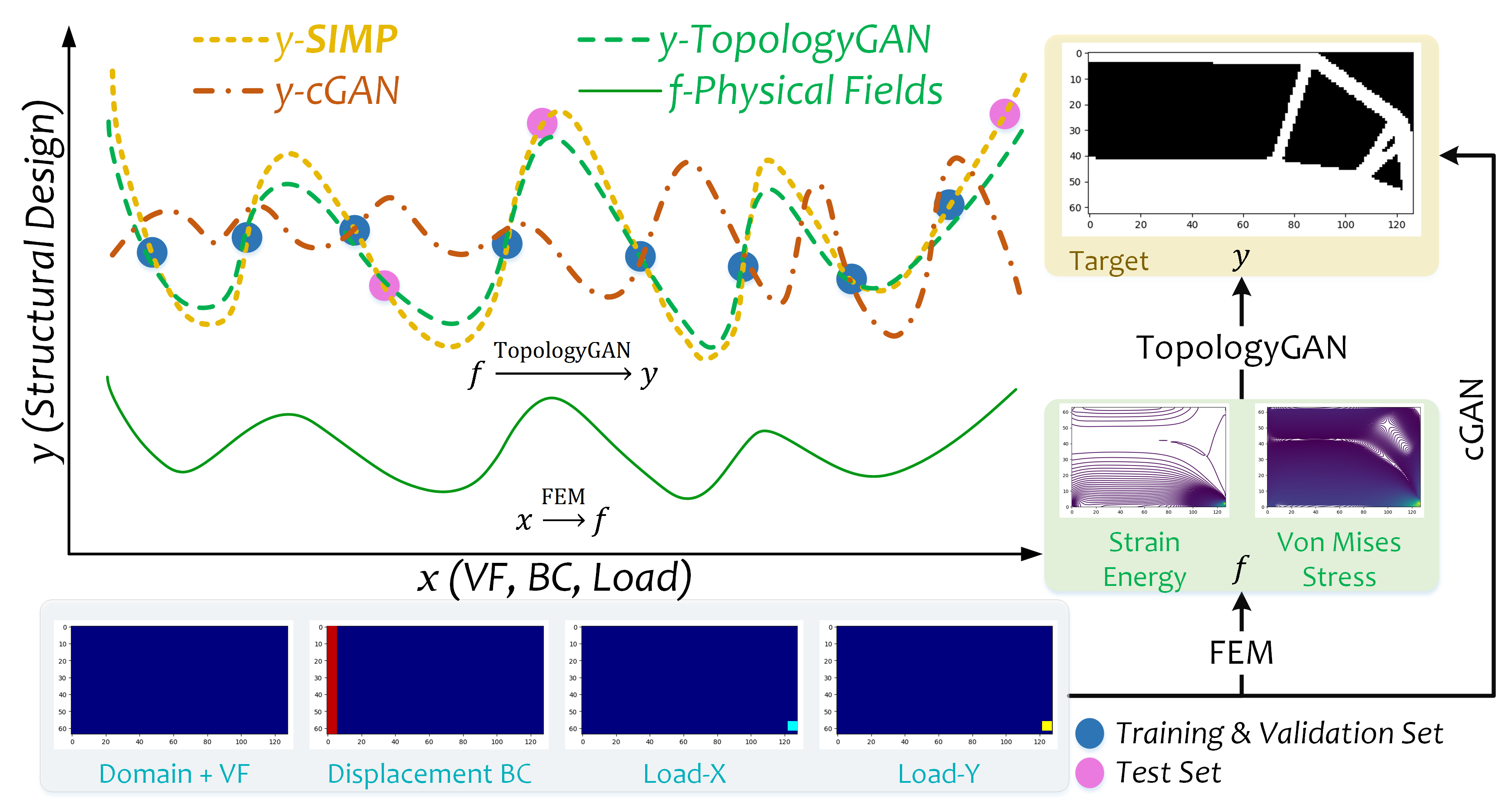}
    \caption{Schematic diagram of TopologyGAN.}
    \label{fig:schematics_topgan}
\end{figure}

As one step toward overcoming this challenge, we propose a new model called TopologyGAN. The method is based on the use of various physical fields such as the strain energy field, von Mises stress field, displacement fields, and strain fields computed on the \emph{initial} (unoptimized) design domain as a way to augment the baseline cGAN with this extra information. In this work, we denote these as the \emph{initial fields} $f$. An illustrative schematic of how TopologyGAN works is shown in Figure \ref{fig:schematics_topgan}. The horizontal axis is composed of the problem input matrices encoding the desired VF, displacement BC, and load. The vertical axis is the resulting structure where three structural designs are generated respectively by the SIMP method as the ground truth, by the cGAN as the baseline, and by TopologyGAN as the proposed model. The baseline cGAN directly maps the inputs to the output structure: $x \xrightarrow{\text{cGAN}} y$. Our proposed TopologyGAN, on the other hand, builds a mapping from the inputs $x$ to the initial fields $f$, followed by a mapping from $f$ to the output structure $y$ as follows: $x \xrightarrow{\text{FEM}} f \xrightarrow{\text{TopologyGAN}} y$. Note that during run time, the initial fields are computed only once. 

Our hypothesis in utilizing such initial fields is that they provide useful information regarding the physical state of the original domain under the inputs $x$ that can be further exploited by the network. For instance, as shown in Figure \ref{fig:schematics_topgan}, both the initial strain energy density and the von Mises stress maps produce scalar fields that are richer in information compared to the original problem input matrices $x$ alone. In particular, the initial fields are hypothesized to produce information $f$ (green solid curve) that correlates well with the final structure $y$ (yellow dashed curve), thereby making the network's remaining task less daunting compared to the scenario of mapping $x$ to $y$ directly. As will be shown in Section \ref{ModelEvaluation}, the initial fields indeed help TopologyGAN attain significantly higher training and test accuracies over the baseline cGAN. 

%From the two shown physical fields of the strain energy density and the von Mises stress, the distribution of the values of the physical fields matrices usually presents the visual similarity with those of the final structural design variables. Such a visual similarity ensures that TopologyGAN more easily extracts the hidden information in the latent space and performs better on unseen BC in the test than cGAN. As expressed vividly in \ref{fig:schematics_topgan}, curves of the physical fields ($f-Physical Fields$), the TopologyGAN output ($y-TopologyGAN$), and the ground truth ($y-SIMP$) have a consistent trend of change with the input design variables. TopologyGAN can coincide with the ground truth better than cGAN, especially for the test set.

\subsection{Network Architecture of TopologyGAN}

\begin{figure}[!ht]
    \centering
    \includegraphics[width=1.0\linewidth]{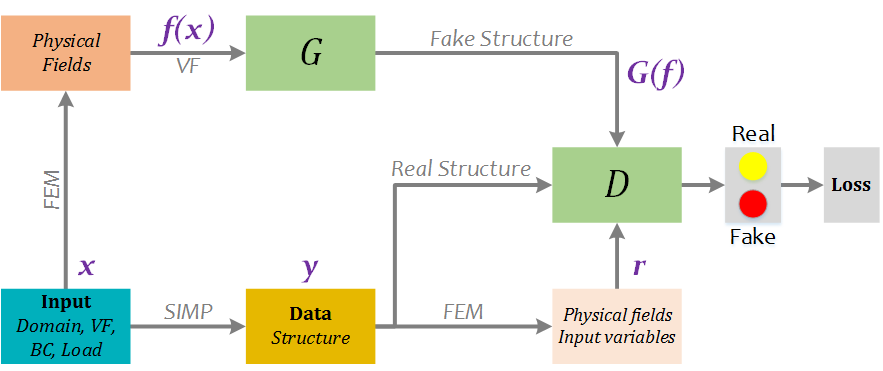}
    \caption{TopologyGAN approach to topology optimization.}
    \label{fig:topologygan}
\end{figure}

TopologyGAN is based on the conditional Pixel2Pixel GAN architecture \cite{isola2017image} and incorporates the FEM and the SIMP solver, as shown in Figure \ref{fig:topologygan}. The generator of TopologyGAN establishes a mapping from the initial physical fields $f$ to the estimated optimized topology $\hat{y}=G(f(x),\rm{VF})$. The ground truth output $y$ is generated by the SIMP method, as was the case with the baseline cGAN shown in Figure \ref{fig:cgan}. In TopoloyGAN, both the problem inputs $x$ and the initial fields $f(x)$ are used as the condition: $r(x)=[x,f(x)]$.

%Here, the CNN block is defined as a serial module of a convolution layer, a batch normalization layer and a rectified linear units (ReLU) layer. By given any specific input variable $x$, the ground truth topological structure $y$ is obtained by the SIMP method. Meanwhile, the initial physical fields $f(x)$ are calculated by FEM and it is used as the input for the generator $G$. Then, the generator $G$ is trained to mimic the optimal structure $\hat{y}$ and to deceive the discriminator $D$. At the same time, the discriminator $D$ is trained to distinguish between the generator's fake structure $\hat{y}$ and the ground truth $y$. 

\begin{figure}[!ht]
    \centering
    \includegraphics[width=1.0\linewidth]{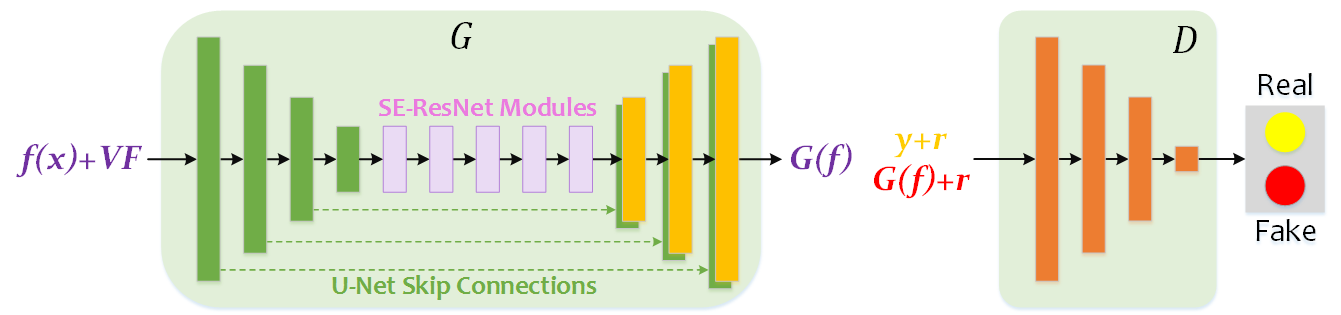}
    \caption{Architectures of the generator $G$ and discriminator $D$.}
    \label{fig:g_and_d}
\end{figure}

Both the generator and the discriminator\ utilize CNN blocks as shown in Figure \ref{fig:g_and_d}. For the generator, we propose a hybrid architecture called U-SE-ResNet. U-SE-ResNet is a U-Net with the SE-ResNet module in a downsampling-upsampling structure. In the upsampling, transposed convolution layers are used instead of pooling layers. According to \cite{isola2017image}, PatchGAN discriminator is adopted because the problem of blurry images caused by failures at high-frequency features such as edges and textures can be alleviated by restricting the GAN discriminator to only model high frequencies. The Sigmoid function is used in the generator's last layer for outputting a grayscale structure $\hat{y}$. 

\subsection{TopologyGAN Loss Function}
As shown in Eq.(\ref{equ:tgan_loss_cgan}) - Eq.(\ref{equ:tgan_loss_vf}), the loss consists of three parts: (1) the loss of TopologyGAN: $\mathcal{L}_\mathit{G,D}^{\mathit{TGAN}}$, (2) the L2 loss of the generator: $\mathcal{L}_{L2}(\mathit{G})$, and (3) the absolute error of the volume fraction of the generator: $\textrm{AE}^{\textrm{VF}}$.

\begin{equation}\label{equ:tgan_loss_cgan}
\begin{split}
\mathcal{L}_\mathit{G,D}^{\mathit{TGAN}}= \mathbb{E}_{(x,y) \sim p_{data}(x,y)}[\log\mathit{D}(r(x),y)]+ \\ \mathbb{E}_{x \sim p_{data}(x)}[\log(1-\mathit{D}(r(x),\mathit{G}(f(x))))]
\end{split}
\end{equation}

\begin{equation}\label{equ:tgan_loss_cgan_g}
\mathcal{L}_\mathit{G}^{\mathit{TGAN}}= \mathbb{E}_{x \sim p_{data}(x)}[\log(1-\mathit{D}(r(x),\mathit{G}(f(x))))]
\end{equation}

\begin{equation}\label{equ:tgan_loss_l2}
\mathcal{L}_{L2}(\mathit{G}) = \mathbb{E}_{x,y}[\left\|y-G(f(x))\right\|_2]
\end{equation}

\begin{equation}\label{equ:tgan_loss_vf}
\textrm{AE}^{\textrm{VF}}_{\mathit{G}} = \mid \textrm{VF}-\widehat{\textrm{VF}} \mid = \frac{1}{N} \mid \sum_{e=1}^{N} (y_e - G(f(x))_e) \mid 
\end{equation}

\begin{equation}\label{equ:tgan_loss_final}
\mathit{G}^* = \text{arg} \mathop{\text{max}}\limits_{\mathit{D}}\mathop{\text{min}}\limits_{\mathit{G}} \ \mathcal{L}_\mathit{G,D}^{\mathit{TGAN}} + \lambda_{1} \mathcal{L}_{L2}(\mathit{G}) + \lambda_{2} \textrm{AE}^{\textrm{VF}}_{\mathit{G}}
\end{equation}

\noindent where the generator $\mathit{G}$ tries to minimize the combined loss $\mathcal{L}_\mathit{G}^{\mathit{TGAN}} + \lambda_{1} \mathcal{L}_{L2}(\mathit{G}) + \lambda_{2} \textrm{AE}^{\textrm{VF}}_{\mathit{G}}$. At the same time, the adversarial $\mathit{D}$ tries to maximize $\mathcal{L}_\mathit{G,D}^{\mathit{TGAN}}$. $r(x)$ is the condition that contains the problem inputs $x$ and the initial physical fields $f(x)$. $n$ is pixel count of the output image. Scalars $\lambda_{1}$ and $\lambda_{2}$ are used to balance the three loss functions. The two scalars are determined empirically using parametric studies. In this paper, $\lambda_{1}=10{,}000$, $\lambda_{2}=1$.

We select the cGAN as our baseline model. Its structure is shown in Figure \ref{fig:cgan}. Its loss is shown in Eq.(\ref{equ:cgan_loss_final}).

%%%%%%%%%%%%%%%%%%%%%%%%%%%%%%%%%%%%%%%%%%%%%%%%%%%%%%%
\section{Experiments}
The proposed TopologyGAN and the baseline cGAN are trained on the same data set that is generated by the SIMP method and then are evaluated and compared on their prediction performance.

\subsection{Experiment Design}
Our code is written in TensorFlow and trained on a NVIDIA GeForce GTX 2080Ti GPU. Adam \cite{kingma2014adam} is used for optimization, combining the advantages of AdaGrad \cite{duchi2011adaptive} and RMSProp \cite{hinton2012neural}.

In addition to a comparison of TopologyGAN and the baseline cGAN, we also study the impact of different physical fields as inputs. Additionally, we compare different generator structures, namely U-Net, SE-ResNet, and the proposed U-SE-ResNet.

\subsection{Data set}
A SIMP-based topology optimization framework called ToPy \cite{hunter2017topy} is used to generate the data set. The 2D domain consists of a $64 \times 128$ grid structure consisting of square elements. A total of $49,078$ optimized structures are generated using ToPy with the randomized conditions as follows:

\begin{description}
\item[$\bullet$] volume fraction: $\left[0.3:0.02:0.5\right]$
\item[$\bullet$] displacement BCs: 42 scenarios
\item[$\bullet$] load position: any point on the boundary of the domain
\item[$\bullet$] load direction: $\left[0:\frac{\pi}{6}:\pi\right]$
\item[$\bullet$] SIMP penalty: 2
\item[$\bullet$] SIMP filter radius: 1.5
\end{description}

The data set is divided into training, validation, and test sets as follows: 4 displacement BCs of the total 42 are randomly selected as the test set. All data samples from the remaining 38 displacement boundary conditions are shuffled and split into training (80\%) and validation (20\%) sets. Note that the boundary conditions in the test set will not have been seen by the network during training.

\begin{figure}[!ht]
    \centering
    \includegraphics[width=1.0\linewidth]{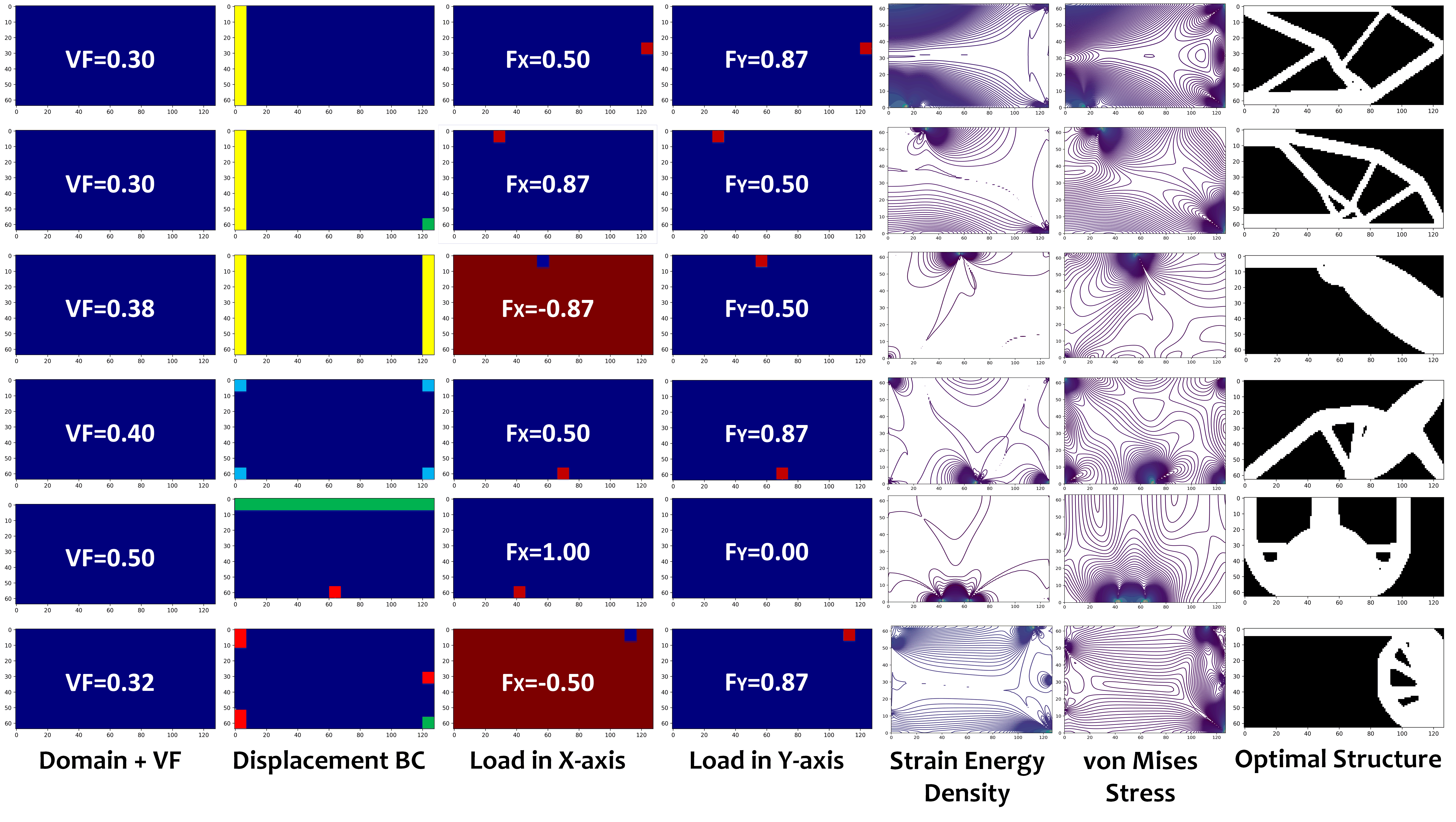}
    \caption{Data samples generated by FEM solver and SIMP solver. Images in each row form one sample. The first four images in each row are input variables $x(\textrm{VF,BC,Load})$, the last image is output $y$. Strain energy density and von Mises stress are shown in level sets with contour lines.}
    \label{fig:data_sample}
\end{figure}

Six samples are randomly selected from the entire data set for an illustration of the input and outputs in Figure \ref{fig:data_sample}. Images in each row form a sample for TopologyGAN. The first four images in each row are input variables $x(\textrm{VF,BCs})$ including volume fraction, displacement boundary condition and load boundary conditions along the x-axis and the y-axis. VF is input as a 2D matrix that has the same dimension as the design domain. The value of each element in the matrix is equal to the VF value. Displacement BCs are represented as a 2D matrix by assigning one of the four integers to each element: $0$ represents unconstrained, $1$ represents $u_x=0$, $2$ represents $u_y=0$, and $3$ represents $u_x=u_y=0$. Based on the input variables, the initial stress ($\sigma=[\sigma_{11},\sigma_{22},\sigma_{12}]$), strain ($\varepsilon=[\varepsilon_{11},\varepsilon_{22},\varepsilon_{12}]$), and displacement ($u=[u_x,u_y]$) fields are computed by FEM. Strain energy density and von Mises stress are shown as the example physical fields, which are expressed in Eq(\ref{equ:vm_stress}) and Eq(\ref{equ:strain_energy}). The last image is the output structure $y$ from SIMP. 

\begin{equation}\label{equ:vm_stress}
\sigma_{\textrm{vm}} = \sqrt{\sigma_{11}^2-\sigma_{11}\sigma_{22}+\sigma_{22}^2+3\sigma_{12}^2}
\end{equation}

\begin{equation}\label{equ:strain_energy}
W = \frac{1}{2} (\sigma_{11} \varepsilon_{11}+\sigma_{22} \varepsilon_{22}+2\sigma_{12} \varepsilon_{12})
\end{equation}

\subsection{Evaluation Metrics}
To assess the performance of TopologyGAN, we use the following four metrics: mean absolute error ($\textrm{MAE}$), mean squared error ($\textrm{MSE}$), relative error of volume fraction ( $\textrm{AE}^{\textrm{VF}}$ ) and its absolute value ( $\mid \textrm{AE}^{\textrm{VF}} \mid$ ), relative error of the compliance ( $\textrm{RE}^{\textrm{C}}$ ) and its absolute value ( $\mid \textrm{RE}^{\textrm{C}}\mid$ ). Denote $\hat{y}$ as the prediction from the generator $G$:

\begin{equation}\label{equ:y_G(x)}
\hat{y} = G(f(x)),
\end{equation}

\noindent and then both the ground truth $y$ and the prediction $\hat{y}$ are reshaped into vectors, with the length of $8{,}192$, from 2D matrices, with the size of $64 \times 128$.

MAE shown in Eq.(\ref{equ:tgan_metric_mae}) and MSE shown in Eq.(\ref{equ:tgan_metric_mse}) are used to evaluate the model. MAE measures the average magnitude of the absolute differences between the prediction values and the ground truth. MSE measures the average squared difference between the estimated values and the ground truth.

\begin{equation}\label{equ:tgan_metric_mae}
\textrm{MAE} = \frac{1}{M} \sum_{i=1}^{M} \mid y^{(i)} - \hat{y}^{(i)} \mid = \frac{1}{M} \sum_{i=1}^{M} \frac{1}{N} \sum_{e=1}^{N} \mid y^{(i)}_e - \hat{y}^{(i)}_e \mid
\end{equation}

\begin{equation}\label{equ:tgan_metric_mse}
\textrm{MSE} = \frac{1}{M} \sum_{i=1}^{M} (y^{(i)} - \hat{y}^{(i)})^2 = \frac{1}{M} \sum_{i=1}^{M} \frac{1}{N} \sum_{e=1}^{N} (y^{(i)}_e - \hat{y}^{(i)}_e)^2
\end{equation}

\noindent where $M$ is the total number of data samples, and $N=8{,}192$ is the number of grid elements.

%$\textrm{R}^2$ is another metric we use to evaluate the trained model. It shows how well our model does compare to a model that simply averages values.

%$\textrm{R}^2$ is theoretically going to be between $-\infty$ and $1$ in and normally ranges from $0$ to $0$. A $\textrm{R}^2$ value close to $1$ means the model error tends to $0$.

%\begin{equation}\label{equ:tgan_metric_r2}
%\textrm{R}^2 = 1-\frac{\sum_{i=1}^{M} (y^{(i)} - %\bar{y})^2}{\sum_{i=1}^{M}(y^{(i)} - %\hat{y}^{(i)})^2}=1-\frac{\sum_{i=1}^{M}(y^{(i)} - %\bar{y})^{\mathsf{T}}(y^{(i)} - \bar{y})}{\sum_{i=1}^{M}(y^{(i)} - %\hat{y}^{(i)})^{\mathsf{T}}(y^{(i)} - \hat{y}^{(i)})}
%\end{equation}

%\noindent where $\bar{y} = \frac{1}{M}\sum_{i=1}^{M} y^{(i)}$ is the mean value of $y$, the superscript $\mathsf{T}$ denotes the transpose operation.
In addition to these commonly used metrics, we define two other metrics for evaluation: (1) ${\textrm{RE}}^{\textrm{VF}}$ is the relative error of the volume fraction between prediction output and ground truth output, and (2) $\textrm{RE}^{\textrm{C}}$ is the ratio between the compliance of the predicted structure and the ground truth structure. These are defined as follows:

%It is necessary to note that $\textrm{RE}^{\textrm{VF}}$ and $\textrm{RE}^{\textrm{C}}$ are statistical variables for each sample and need further statistical analysis on a mass of data samples in distribution charts or histograms.

\begin{equation}\label{equ:vf}
\textrm{VF} = \frac{1}{N} \sum_{e=1}^{N} y_e
\end{equation}

\begin{equation}\label{equ:tgan_metric_vf}
\textrm{RE}^{\textrm{VF}} = \frac{\widehat{\textrm{VF}}-\textrm{VF}}{\textrm{VF}} = \frac{\sum_{e=1}^{N} (\hat{y}_e-y_e)}{\sum_{e=1}^{N} y_e}
\end{equation}

\begin{equation}\label{equ:compliance}
C(y) = \mathbf{U}^T \mathbf{KU}=\sum_{e=1}^{N} (y_e)^p \mathbf{u}_e^T \mathbf{k}_0 \mathbf{u}_e
\end{equation}

\begin{equation}\label{equ:tgan_metric_compliance}
\textrm{RE}^{\textrm{C}} = \frac{C(\hat{y})-C(y)}{C(y)}
\end{equation}

\noindent where $C(y)$ is the compliance of the predicted structure under the given loads.

%%%%%%%%%%%%%%%%%%%%%%%%%%%%%%%%%%%%%%%%%%%%%%%%%%%%%%%
\section{Results and Discussions}
We find that TopologyGAN outperforms the baseline cGAN in both the training and test. Furthermore, we analyze the physical field selection and the network architecture for the generator. We find that the VF+$\sigma_{\text{vm}}$+$W$ combination performs significantly better than all other combinations. Moreover, to find the best generator structure, we compare the U-SE-ResNet, U-Net, and SE-ResNet. A comparative experiment is conducted using different generators on the same discriminator and the data set. The results show that the proposed U-SE-ResNet outperforms the other two structures. 

\subsection{Model Evaluation}
\label{ModelEvaluation}
Our training results of TopologyGAN and the baseline cGAN are summarized in Table \ref{tab:comparison_two}. It can be seen that TopoloyGAN achieves a $3\times$ lower MSE error and $2.5\times$ lower MAE error than that of the cGAN. More surprisingly, both the MSE and MAE of TopologyGAN on the \emph{test} set are lower than those of cGAN on the \emph{training} set.

\begin{figure}[!ht]
    \centering
    \includegraphics[width=0.7\linewidth]{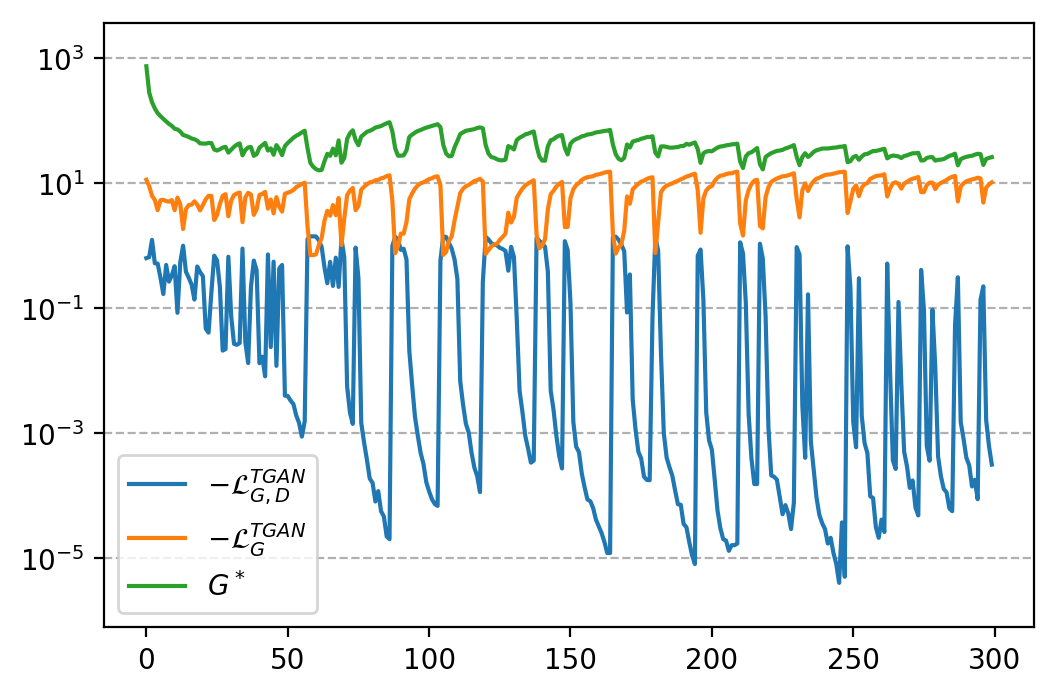}
    \caption{Loss functions of TGAN, Discriminator and Generator in training.}
    \label{fig:tgan_loss}
\end{figure}

Loss functions of TopologyGAN are shown in Figure \ref{fig:tgan_loss}, where $\mathcal{L}_\mathit{G, D}^{\mathit{TGAN}}$ is the discriminator loss of TopologyGAN, $\mathcal{L}_\mathit{G}^{\mathit{TGAN}}$ is the generator loss of TopologyGAN, and $\mathit{G^*}$ is the whole objective of the generator. As the training progresses, it can be seen that $\mathcal{L}_\mathit{G}^{\mathit{TGAN}}$ and $\mathit{G^*}$ decrease gradually, and $\mathcal{L}_\mathit{G, D}^{\mathit{TGAN}}$ oscillates and tends to balance.

\begin{table*}[!ht]
    \caption{Comparison of results between TopologyGAN and cGAN.}
    %\begin{center}
    \centering
    \begin{tabular}{c ccc ccc}
        \toprule
        \multirow{2}{*}{Model} & \multicolumn{3}{c}{MAE} & \multicolumn{3}{c}{MSE} \\
        \cmidrule(r){2-4} \cmidrule(r){5-7}
            & Training & Validation & Test & Training & Validation & Test \\
        \midrule
        TopologyGAN & 0.001808 & 0.019133 & 0.070128 & 0.001340 & 0.018022 & 0.059943\\
        Baseline: cGAN & 0.088257 & 0.100565 & 0.181095 & 0.085916 & 0.097966 & 0.175226\\
        \bottomrule
    \end{tabular}
    %\end{center}
    \label{tab:comparison_two}
\end{table*}

\begin{figure}[!ht]
    \centering
    \includegraphics[width=0.7\linewidth]{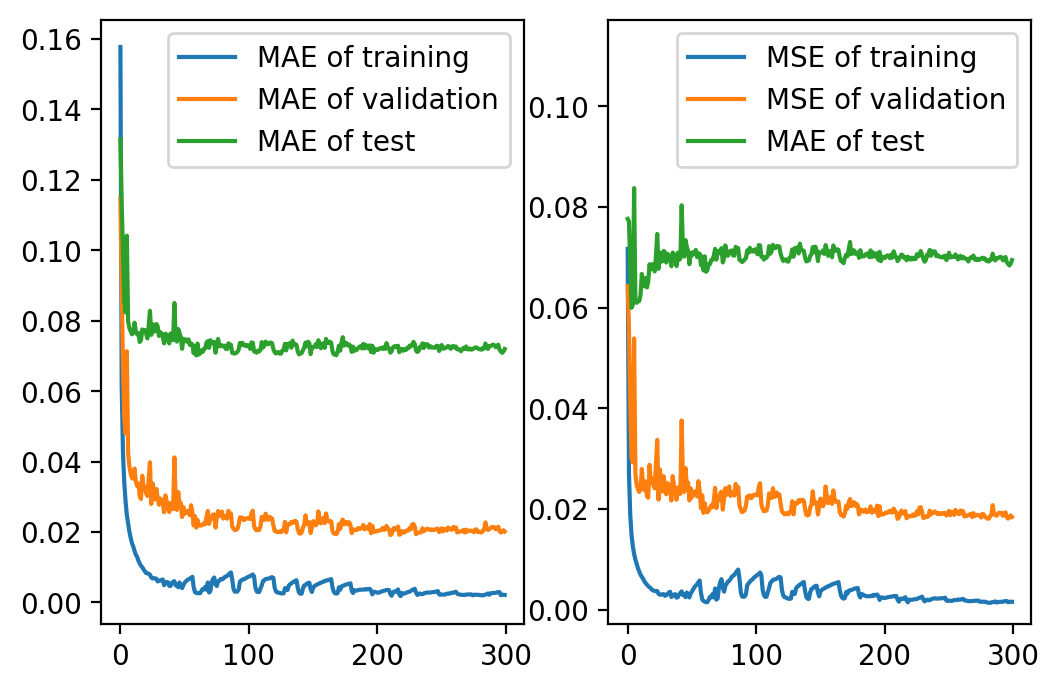}
    \caption{MAE and MSE of TopologyGAN in training.}
    \label{fig:tgan_mae_mse}
\end{figure}

\subsection{Accuracy and Performance}
\begin{figure}[!ht]
    \centering
    \includegraphics[width=1.0\linewidth]{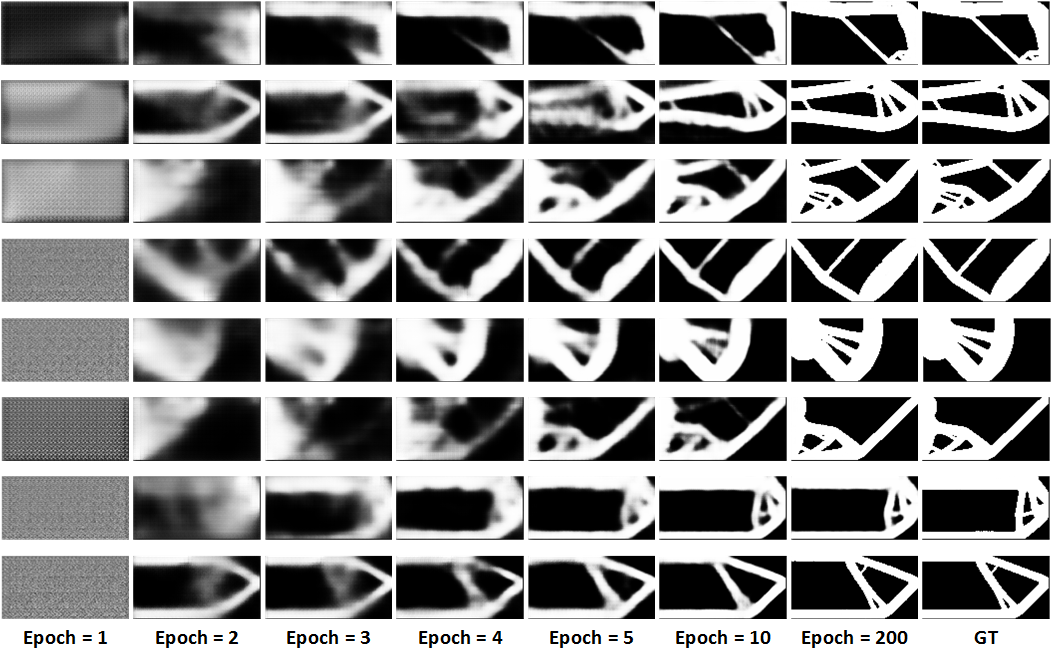}
    \caption{Evolution of TopologyGAN predictions during training.}
    \label{fig:evolutionary_output}
\end{figure}

The accuracy and performance of the trained TopologyGAN are discussed to show the prediction performance. As shown in Figure \ref{fig:evolutionary_output}, the generated structures from TopologyGAN become increasingly more defined over the training epochs. Each row denotes a randomly selected sample from the training set. After two-hundred epochs, the predictions become virtually indistinguishable from the ground truth (GT) structures.

\begin{figure}[!ht]
    \centering
    \includegraphics[width=1.0\linewidth]{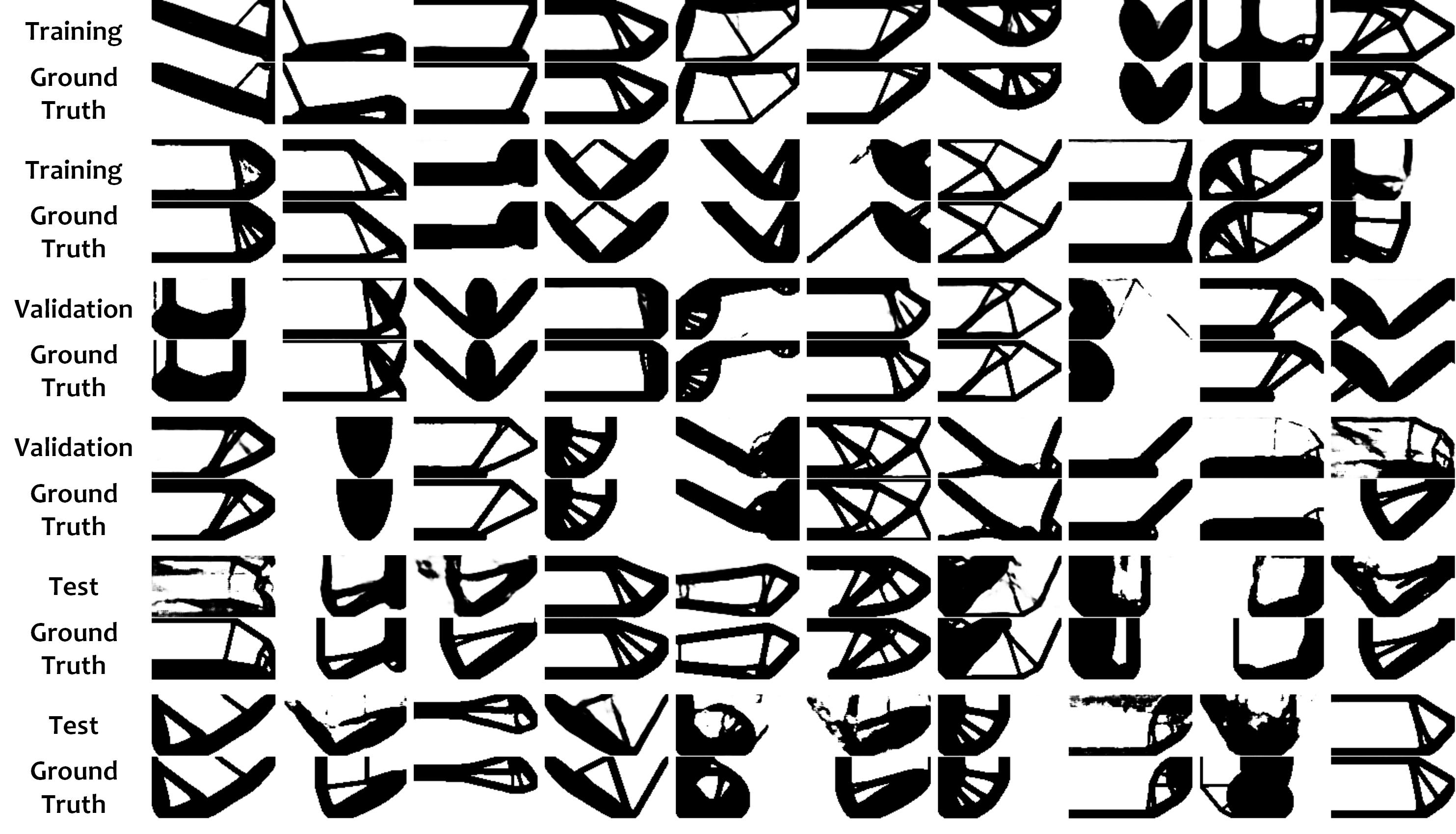}
    \caption{Comparison of the predictions of the fully trained TopoloyGAN on training, validation and test sets.}
    \label{fig:topology_output}
\end{figure}

To visually compare the prediction accuracy of the fully trained TopologyGAN on the training, validation, and test sets, Figure \ref{fig:topology_output} shows the computed results and the corresponding ground truth structures. By comparing each set of images, we find that TopologyGAN performs well on the training and validation sets. The performance of TopologyGAN on the test set is expectedly lower.

To further quantify the performance, we compute the VF and the compliance of the resulting structures and compare them to those of the ground truth structures. Based on the fully trained TopologyGAN, we randomly select 640 samples from each of the training and test sets. For each data sample, this results in two structures: a prediction $\hat{y}$ and a ground truth $y$. The relative error of the volume fraction $\textrm{RE}^{\textrm{VF}}$ and the relative error of compliance $\textrm{RE}^{\textrm{C}}$ are computed respectively using Eq.(\ref{equ:tgan_metric_vf}) and Eq.(\ref{equ:tgan_metric_compliance}).

\begin{figure}[!ht]
    \centering
    \includegraphics[width=1.0\linewidth]{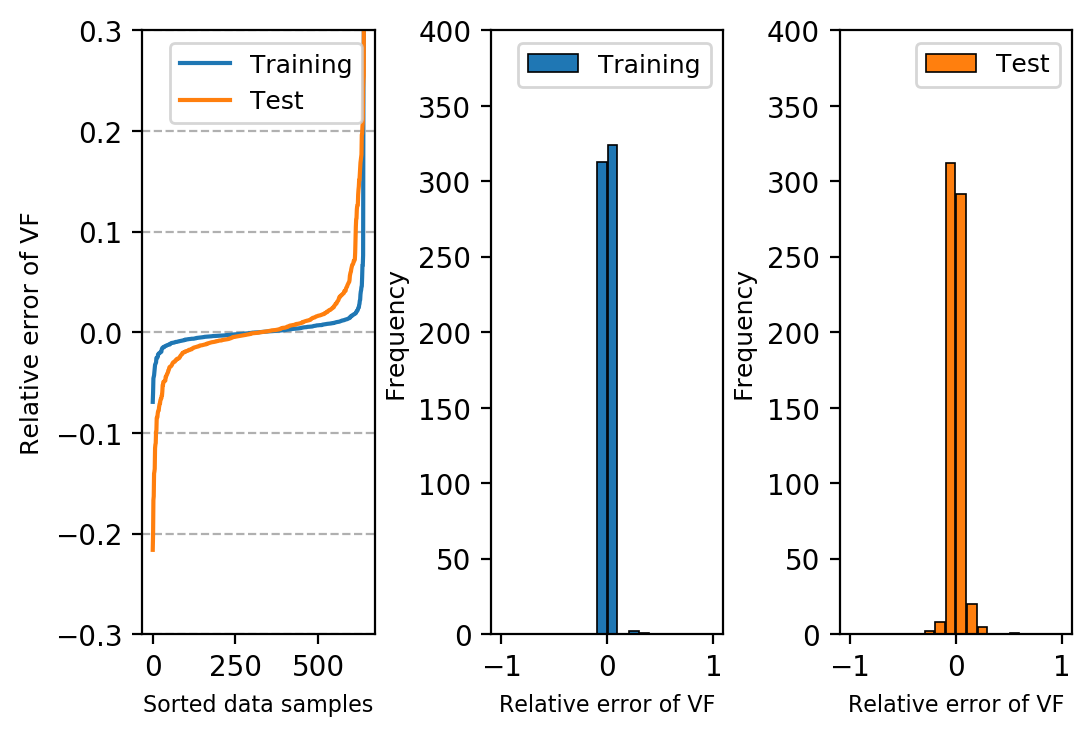}
    \caption{Comparison of volume fraction on training and test sets}
    \label{fig:vf_comparison}
\end{figure}

\begin{figure}[!ht]
    \centering
    \includegraphics[width=1.0\linewidth]{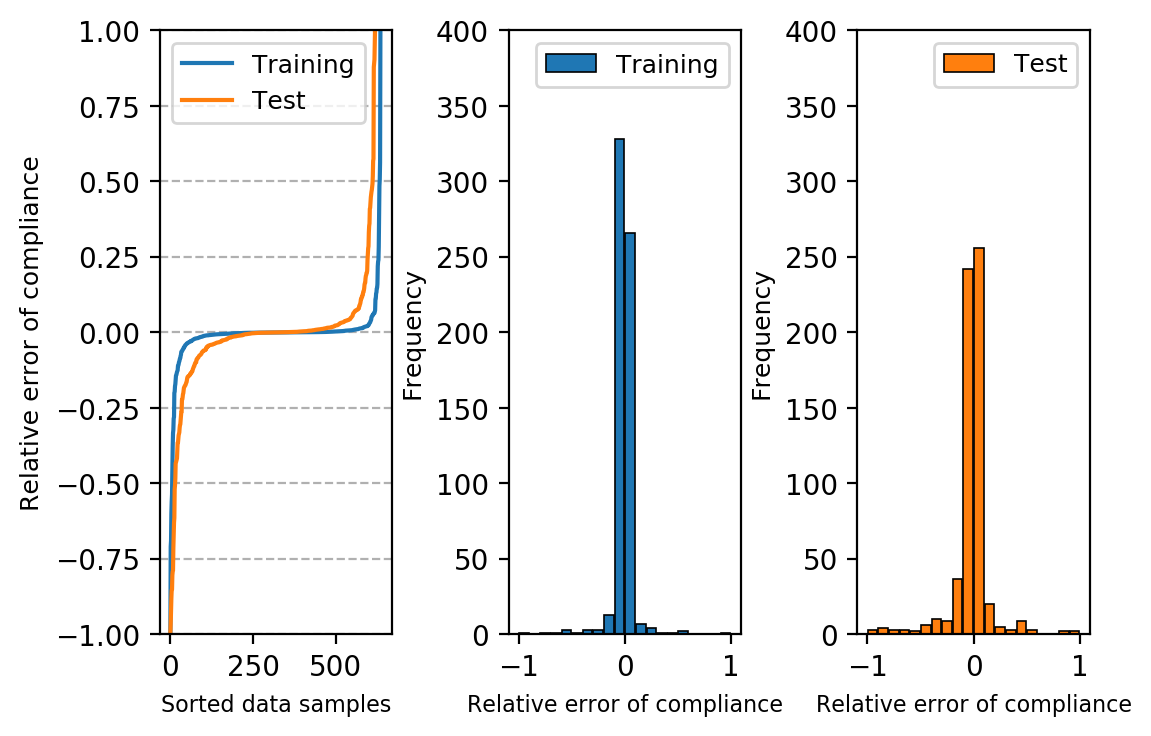}
    \caption{Comparison of compliance on training and test sets.}
    \label{fig:compliance_comparison}
\end{figure}

The sorted ${\textrm{RE}}^{\textrm{VF}}$ values for the training and test sets are shown in the first diagram of Figure \ref{fig:vf_comparison}. Histograms of ${\textrm{RE}}^{\textrm{VF}}$ in the training and the test are also shown in Figure \ref{fig:vf_comparison}. ${\textrm{RE}}^{\textrm{VF}}$ is close to zero for the majority of the samples. Additionally, the histogram of ${\textrm{RE}}^{\textrm{VF}}$ on the training set is more concentrated around zero than that on the test set as expected. ${\textrm{RE}}^{\textrm{C}}$ shows as similar trend as shown in \ref{fig:compliance_comparison}. 

\subsection{Comparison and Selection of Physical Fields}
In addition to the combination of von Mises stress $\sigma_{\text{vm}}$ and strain energy density $W$, various physical field combinations are studied as the inputs. The comparison is shown in Table \ref{tab:physical_fields}. The results indicate that \textnumero 8 - $\left[ \text{VF}+U+\sigma_{\text{vm}}+W \right] $ has the best performance on the training and validation sets, but \textnumero 4 - $\left[ \text{VF}+\sigma_{\text{vm}}+W \right]$ shows a better prediction performance on the test set.

\begin{table*}[!ht]
    \caption{Comparison and selection of physical fields as GAN inputs.}
    \centering
    \begin{tabular}{cl ccc ccc}
        \toprule
        \multicolumn{2}{c}{Metrics} & \multicolumn{3}{c}{MAE} & \multicolumn{3}{c}{MSE}\\
        \cmidrule(r){1-2} \cmidrule(r){3-5} \cmidrule(r){6-8}
        \textnumero & Physical Fields & Training & Validation & Test & Training & Validation & Test\\
        \midrule
        0 & Baseline & 0.088257 & 0.100565 & 0.181095 & 0.085916 & 0.097966 & 0.175226\\
        1 & VF+$U$ & 0.001941 & 0.039551 & 0.105560 & 0.002391 & 0.030464 & 0.099863\\
        2 & VF+$W$ & 0.001781 & 0.039145 & 0.100626 & 0.002343 & 0.033687 & 0.094903\\
        3 & VF+$\sigma_{\text{vm}}$ &0.001758 & 0.040411 & 0.098619 & 0.002333 & 0.035467 & 0.081702\\
        4 & VF+$\sigma_{\text{vm}}$+$W$ & 0.001808 & \cellcolor{yellow!40}0.019133 & \cellcolor{yellow!40}0.070128 & 0.001340 & \cellcolor{yellow!40}0.018022 & \cellcolor{yellow!40}0.059943\\
        5 & VF+$\sigma$ & 0.002339 & 0.037105 & 0.101626 & 0.002382 & 0.031802 & 0.095526\\
        6 & VF+$\varepsilon$ & 0.001729 & 0.034620 & 0.093073 & 0.002306 & 0.029235 & 0.087086\\
        7 & VF+$\sigma$+$\varepsilon$ & 0.010823 & 0.037518 & 0.122126 & 0.001976 & 0.032496 & 0.092162\\
        8 & VF+$U$+$\sigma_{\text{vm}}$+$W$ & \cellcolor{yellow!40}0.000942 & 0.019519 & 0.088748 & \cellcolor{yellow!40}0.001099 & 0.031628 & 0.079184\\
        9 & VF+LP & 0.001914 & 0.033079 & 0.139781 & 0.00328 & 0.037652 & 0.100326\\
        \midrule
        \multicolumn{8}{p{13.5cm}}{\raggedright Note: VF is volume fraction, $F$ is external loads, $U$ is displacement field, $\sigma_{\text{vm}}$ is von Mises stress, $W$ is strain energy density, $\sigma$ is stress field, $\varepsilon$ is strain field, and LP is load path vector. \colorbox{yellow!40}{Number} marked in yellow is the minimum in each column.}\\
        \bottomrule
      \end{tabular}
      \label{tab:physical_fields}
\end{table*}

\subsection{Generator Architecture} 
\begin{figure}[!ht]
    \centering
    \includegraphics[width=1.0\linewidth]{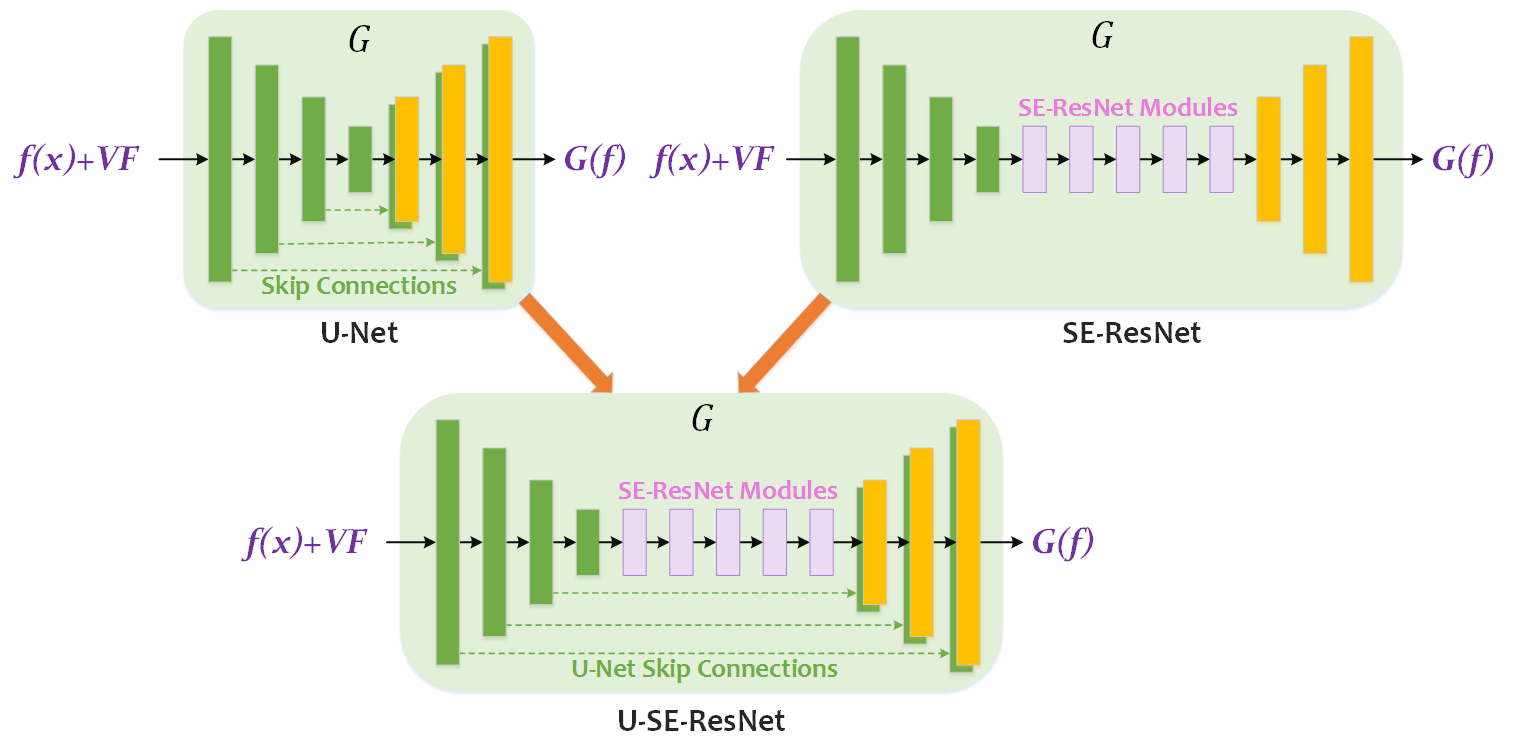}
    \caption{Architectures of U-Net, SE-ResNet, and U-SE-ResNet}
    \label{fig:g_u_resnet}
\end{figure}

The most important feature of the U-Net is the skip connections between mirrored downsampling and upsampling layers to transfer local information and to merge features at the various resolution levels. The most distinct feature of SE-ResNet is the shortcut connection which performs identity mapping added to the output of the stacked layers to dynamically select the layer depth for the desired underlying mapping. Served as the generator of TopologyGAN, the proposed architecture U-SE-ResNet in this article combines U-Net and SE-ResNet, as shown in Figure \ref{fig:g_u_resnet}. We train the TopologyGAN model with different generator architectures, which are U-Net, SE-ResNet, and U-SE-ResNet. The training results are shown in Table \ref{tab:generator_architecture}. U-SE-ResNet has the best performance of the three. This is due to the fact that the U-SE-ResNet combines the advantages of the U-Net and the SE-ResNet, which improves the model flexibility in the local information transfer and the adjustable network depth.

\begin{table*}[!ht]
    \caption{Training results of the three generator architectures}
    %\begin{center}
    \centering
    \begin{tabular}{c ccc ccc}
        \toprule
        \multirow{2}{*}{Architecture} & \multicolumn{3}{c}{MAE} & \multicolumn{3}{c}{MSE}\\
        \cmidrule(r){2-4} \cmidrule(r){5-7}
            & Training & Validation & Test & Training & Validation & Test\\
        \midrule
        U-Net & 0.002908 & 0.027434 & 0.101455 & 0.002471 & 0.029133 & 0.098439\\
        SE-ResNet & 0.008597 & 0.100675 & 0.142915 & 0.058231 & 0.089755 & 0.157362\\
        U-SE-ResNet & 0.001808 & 0.019133 & 0.070128 & 0.001340 & 0.018022 & 0.059943\\
        \bottomrule
    \end{tabular}
    %\end{center}
    \label{tab:generator_architecture}
\end{table*}

\subsection{Limitations and Future Work}
TopologyGAN exhibits good performance compared to the baseline cGAN and generalizes to previously unseen boundary conditions. However, there are several limitations. First, there is no in-built mechanism that ensures a single connected component (or the avoidance of the checkerboard pattern) outside of the penalization enhanced SIMP-based ground truth training data the network observes. Second, while VF is typically employed as an upper bound for the amount of material that SIMP can utilize, TopologyGAN treats VF as a condition to match as closely as possible (minimizing the absolute error of the target and generated VF) rather than a true inequality constraint. Third, TopologyGAN is implemented for 2D topology optimization and, while the approach will be similar in nature, will require modifications to extend to 3D.

TopologyGAN only takes advantage of the fields computed on the original, unoptimized domain. However, theoretically, there is no restriction on which fields can be used to augment the network. In particular, a new network could be devised where the structure generated by TopologyGAN is reassessed to compute the fields of interest and this information can be fed back to the network for improved predictions in several iterative (yet fast) steps. These limitations and observations will be the subject of our future work.

%In future work, we will mainly focus on the second and third limitations. To discourage the checkerboard pattern, we would like to adopt the mesh-independent filter function, which is used in SIMP, as a part of the loss function of the generator. To suppress the disconnected design elements, we will add structural compliance to the loss function of the generator. Moreover, we would like to extend the TopologyGAN model to 3D domains for more practical applications.

%%%%%%%%%%%%%%%%%%%%%%%%%%%%%%%%%%%%%%%%%%%%%%%%%%%%%%%
\section{Conclusions}
We present a new deep learning based generative model called TopologyGAN for topology optimization where the displacement and load boundary conditions, and the target volume fraction are prescribed as inputs. TopologyGAN uses dense initial fields computed over the original, unoptimized domain to augment the conventional inputs. These new fields prove to be helpful for the network to outperform a baseline cGAN in significant ways. In addition, we propose a new hybrid generator architecture called U-SE-ResNet that combines the features of U-Net and SE-ResNet. The use of the initial fields, as well as the U-SE-ResNet model, allows TopologyGAN to significantly reduce the test errors on problems involving previously unseen boundary conditions. 

%% for ASME Conference and Journal
%\bibliographystyle{asmems4}
%\bibliography{references}

%% for arXiv
%% 1. After compile by overleaf, select logs and output files, generate bbl files (such as manuscript.bbl). Or run LatexMk to generate a bbl file.
%% 2.Upload output.bbl file to the arXiv, add \input{references.bbl} at the end of the file before \end{docuement}
\bibliographystyle{unsrt}

%\bibliography{references}
\end{document}